\documentclass[twoside,fleqn]{article}
\usepackage{espcrc2}
\usepackage{theorem}
\usepackage{epsfig}
\usepackage{graphicx}

\theoremstyle{break}    
\theoremstyle{plain}    
\theoremstyle{plain}    
\theoremstyle{plain}    
{\theorembodyfont{\rmfamily}     }
{\theorembodyfont{\rmfamily}     }

\def\gsim{{\mathrel{\raise2pt\hbox to 8pt{\raise -5pt\hbox{$\sim$}\hss{$>$}}}}}
\def\rsim{{\mathrel{\raise2pt\hbox to 8pt{\raise -5pt\hbox{$\sim$}\hss{$>$}}}}}
\def\lsim{{\mathrel{\raise2pt\hbox to 8pt{\raise -5pt\hbox{$\sim$}\hss{$<$}}}}}

\begin{document}

\title{
       \begin{flushright}\normalsize
	    \vskip -0.9 cm
       \end{flushright}
	\vskip -0.4 cm
	\vspace{-0.5cm}
Calculating $\epsilon'/\epsilon$ using HYP staggered fermions\thanks{
Presented by W.~Lee.  Research supported in part by BK21, by
Interdisciplinary Research Grant of Seoul National University, by
KOSEF contract R01-2003-000-10229-0, by US-DOE contracts
DE-FG03-96ER40956, W-7405-ENG-36, KA-04-01010-E161, and NERSC.}  }
\author{T.~Bhattacharya\address[LANL]{MS--B285, T-8,
                Los Alamos National Lab, 
		Los Alamos, New Mexico 87545, USA},
        G.T.~Fleming\address{Jefferson Laboratory,
	        MS 12H2, 12000 Jefferson Avenue, Newport News,
	        VA 23606, USA},
        R.~Gupta\addressmark[LANL],
        G.~Kilcup\address{Department of Physics,
                Ohio State University, Columbus, OH 43210, USA},
        W.~Lee\address{School of Physics, 
		Seoul National University,
		Seoul, 151-747, South Korea}
        and
        S.~Sharpe\address{Department of Physics,
                University of Washington, 
		Seattle, WA 98195, USA}
}
\begin{abstract}
We present preliminary results for $\epsilon'/\epsilon$ calculated
using HYP staggered fermions in the quenched approximation.
We compare different choices of quenched penguin operators.
\end{abstract}

\maketitle

There are a number of advantages to using staggered fermions for
calculating weak matrix elements relevant to CP violation in the
neutral kaon system. They retain sufficient chiral symmetry to protect
the weak operators from mixing with lower dimensional
operators. Simulating them is considerably cheaper
than domain wall and overlap fermions. On the other hand
{\em unimproved} staggered fermions have a number of drawbacks: 1-loop
corrections to operator renormalization are large ($\approx 100$\%),
scaling violations, even though they begin at order $a^2$, are large
as is the breaking of the SU(4) taste symmetry. Some of
these limitations can be alleviated by improving staggered fermions
using fat links.

Explicit 1-loop calculations for bilinear operators show that taste
symmetry breaking and the renormalization corrections can be reduced
by an order of magnitude by improving the lattice action. The greatest
reduction is observed for Fat7 and mean field improved
HYP/$\overline{\rm Fat7}$ \cite{ref:wlee:1}. Similarly, the complete set of
1-loop matching coefficients for the four-fermion operators relevant
to CP violation also have small corrections at $1/a \approx 2$ GeV
\cite{ref:wlee:2,ref:wlee:3}. Lastly, numerical simulations show
that HYP smeared links reduce taste symmetry breaking in the pion
multiplet \cite{ref:hasenfratz:1}. In light of these results, we have chosen
to use
HYP/$\overline{\rm Fat7}$ staggered fermions to calculate matrix
elements relevant to $\epsilon'/\epsilon$.
In the standard model, $\epsilon'/\epsilon$ is: 
\begin{eqnarray}
\epsilon'/\epsilon &=& Im(V_{ts}^* V_{td})
    \Big[ P^{(1/2)} - P^{(3/2)} \Big] 
    \label{eq:e'/e} \\
    P^{(1/2)} &=& r \sum_{i=3}^{10} y_i(\mu) \langle O_i \rangle_0(\mu)
    (1 - \Omega_{\eta+\eta'} ) \\
    P^{(3/2)} &=& \frac{r}{\omega} \sum_{i=7}^{10} y_i(\mu) 
    \langle O_i \rangle_2(\mu) \\
    r &=& \frac{ G_F \omega }{ 2 | \epsilon | Re A_0 }
\end{eqnarray}
where $V_{ij}$ are elements of the CKM matrix, $\omega = 1/22.2$
quantifies the $\Delta I = 1/2$ rule and $\Omega_{\eta+\eta'}=0.060(77)$
represents the isospin breaking effect \cite{ref:pich:0}.
$P^{(1/2)}$ ($P^{(3/2)}$) are the $\Delta I=1/2$ ($\Delta
I=3/2$) contribution to $\epsilon'/\epsilon$ and the 
$y_i(\mu)$ are Wilson coefficients in the operator product expansion,
given in \cite{ref:buras:1}.
The sum over operators $O_i$ includes the QCD ($i=3,4,5,6$) and
electroweak ($i=7,8,9,10$) penguin operators. The current-current
operators ($i=1,2$) do not contribute to CP violation. On the lattice
we calculate the matrix elements 
$\langle \pi | O_i(\mu) | K \rangle$ and $\langle 0 | O_i(\mu) | K
\rangle$ and relate them to $\langle O_i \rangle_I(\mu) \equiv \langle
\pi\pi_I | O_i(\mu) | K \rangle$ using chiral perturbation theory at
the leading order. Here the subscript $I$ refers to the isospin of the
final two pion state.
%

%
%
\begin{figure}[t!]
\epsfig{file=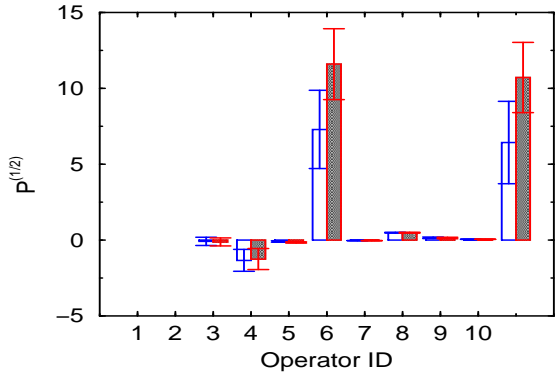, height=12pc, width=17pc}
\vspace*{-6mm}
\caption{Contributions to $P^{(1/2)}$, using linear chiral fits. Empty
(shaded) columns correspond to the STD (GP) methods for $\langle
Q_{5,6} \rangle$.}
\label{fig:p0:std+gp:linear}
\vspace*{-3mm}
\end{figure}

Matrix elements are calculated on 218 quenched gauge
configurations at $\beta = 6.0$ on $16^3 \times 64$ lattices.
Results are converted into the continuum NDR scheme (matching at
$q^*=1/a=1.95$ GeV), evolved to the charm quark mass $m_c = 1.3$ GeV
(using running with $N_f=3$)
\cite{ref:wlee:4} and then combined with the Wilson coefficients
$y_i(m_c)$~\cite{ref:buras:1} to obtain $P^{(1/2)}$ and $P^{(3/2)}$.
We compare two ways of transcribing operators $O_5$ and
$O_6$ to the quenched theory. In the first (the standard [STD] method), the
operators  belong to the singlet representation of $SU(3)_R$,
while in the second (the Golterman-Pallante [GP] method~\cite{GP}) they belong
to the singlet representation of the graded group $SU(3|3)_R$. They
differ because a singlet under $SU(3)_R$ is a combination of
singlet and non-singlet representations of the $SU(3|3)_R$.
Since the dominant contribution to $P^{(1/2)}$ comes from $O_6$, the
difference between these two methods can and, as we show, does have a
significant impact on $\epsilon'/\epsilon$.

A second important issue in the analysis is the function used to
extrapolate the data to the chiral limit. 
First, we take an appropriate ratio to obtain quantities
which does not vanish in the chiral limit. 
We would then like to fit the data to the general form
\begin{equation}
c_1 + c_2 m_K^2 + c_3 m_K^2 \log(m_K^2) + c_4 (m_K^2)^2 \,,
\end{equation}
except that we lack measurements over a sufficient range and number of
masses to simultaneously determine all the $c_i$ parameters.
Thus, we either exclude the logarithm ($c_3=0$: ``quadratic fit"), or
the quartic term ($c_4=0$: ``log fit"), or both ($c_3=c_4=0$: ``linear
fit").
In general, our data cannot distinguish between the fits.
In most cases there is no useful theoretical guidance for the
parameter of the fits, and we compare all three.
Results for $P^{(1/2)}$ using the both STD and GP transcriptions of
$\langle O_5 \rangle$ and $\langle O_6 \rangle$ and linear
extrapolation to the chiral limit are shown in
Fig.~\ref{fig:p0:std+gp:linear}.
The last column in Fig.~\ref{fig:p0:std+gp:linear} is the sum of
individual contributions. The dominant contributions are from
$\langle O_6 \rangle$ and $\langle O_4 \rangle$, which partially
cancel. 
Note that the evolution from $q^*$ to $m_c$ mixes $O_{5,6}$ with 
other operators, so all contributions depend on the choice of the 
operator.
Clearly the choice of operator makes a large difference to the final
result.

%
%
\begin{figure}[t!]
\epsfig{file=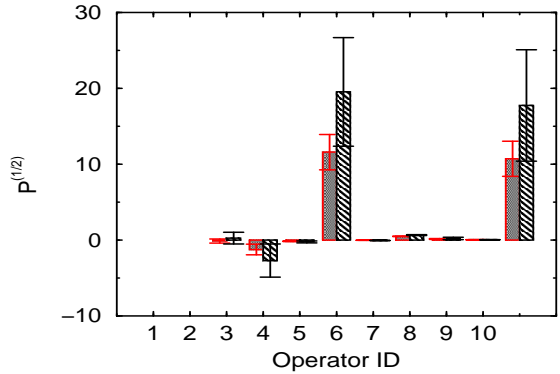, height=12pc, width=17pc}
\vspace*{-6mm}
\caption{Contributions to $P^{(1/2)}$, using linear (left columns)
and quadratic (right columns) chiral fits, and the GP
method for $\langle Q_{5,6} \rangle$.}
\label{fig:p0:GP:lin+quad}
\vspace*{-6mm}
\end{figure}

\begin{figure}[t!]
\epsfig{file=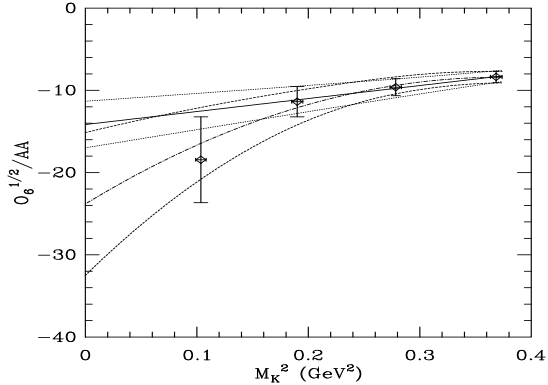, height=12pc, width=17pc}
\vspace*{-6mm}
\caption{Linear and quadratic fits to $\langle \pi | Q_6 | K \rangle /
(m_K^2 f^2)$ (GP method).}
\label{fig:Q6:lin+quad}
\vspace*{-6mm}
\end{figure}

Figure~\ref{fig:p0:GP:lin+quad} illustrates the dependence of the
results on the fitting function used in the chiral extrapolation. The
errors increase on going from linear to quadratic to log fit so much
so that the difference in the result is only about one standard
deviation.  The dominant source of the uncertainty is the value and
error on the point at the lightest quark mass.  This is illustrated in
Fig.~\ref{fig:Q6:lin+quad} which compares linear and quadratic chiral
extrapolations for the ratio $\langle Q_6\rangle/(m_K^2 f^2)$, which should
be non-vanishing in the chiral limit.  The log fit, which is not shown
to maintain clarity, extrapolates to $-30(12)$. So, on the one hand,
one needs masses at least this light ($\approx m_s/5$) to do
reasonable chiral extrapolations, but on the other hand the
statistical errors and possibly finite volume effects grow rapidly
with $1/m$.

%
%

\begin{figure}[t!]
\epsfig{file=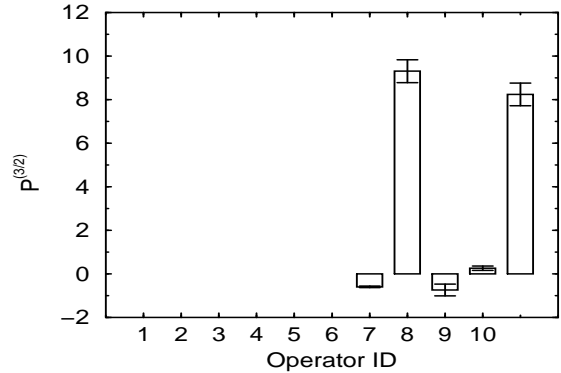, height=12pc, width=17pc}
\vspace*{-6mm}
\caption{Contributions to $P^{(3/2)}$.}
\label{fig:p2}
\vspace*{-6mm}
\end{figure}

Lastly, in Fig.~\ref{fig:p2}, we show the contributions to $P^{(3/2)}$
using linear fits. Here there are no chiral logarithms in the quenched
theory~\cite{GP} and there is no ambiguity in the operator definition.
The contribution from $\langle O_8 \rangle$ dominates, as expected
from a large $N_c$ analysis.

Using Eq.~(\ref{eq:e'/e}) our preliminary estimates for $\epsilon'/\epsilon$ 
for the STD operators are 
  \[
  \frac{\epsilon'}{\epsilon}\bigg|_{\rm(STD)} = \left\{
  \begin{array}{ll}
    -2.4(34)   \times 10^{-4}  &  \mbox{(linear fit)} \\
    -12.8(114) \times 10^{-4}  &  \mbox{(quad. fit)} \\
    -20.1(175) \times 10^{-4}  &  \mbox{(log. fit)}
  \end{array}
  \right.  \,.
  \]
These results are consistent with previous estimates obtained using
domain wall fermions (and STD operators)
\cite{ref:CP-PACS:0,ref:RBC:0}.  On the other hand the GP method gives
a more positive estimate for $\langle O_6 \rangle$, and consequently
$\epsilon'/\epsilon$:
  \[
    \frac{\epsilon'}{\epsilon}\bigg|_{\rm (GP)} = \left\{
    \begin{array}{ll}
    +3.2(29)   \times 10^{-4}   &  \mbox{(linear fit)} \\
    +8.8(93)   \times 10^{-4}   &  \mbox{(quad. fit)} \\
    +13.1(142) \times 10^{-4}  &  \mbox{(log. fit)}
    \end{array}
    \right.  \,,
  \] 
The very large errors in these results are caused by the cancellation
between the $P^{(1/2)}$ and $P^{(3/2)}$ contributions.  
Nevertheless, it appears that the difference between GP and STD, which
is a 2 sigma effect, is significant.
To clarify this point, and to improve the chiral extrapolations,
we are extending the current run to 400 configurations with 4 additional
quark masses.
Ultimately, of course, one needs to do partially quenched simulations
where the operator ambiguity is less important and the coefficients of
the chiral logarithms are known.
%

%
%
%

%
%

This calculation is being done on the Columbia QCDSP supercomputer.
We thank N.~Christ, C.~Jung, C.~Kim, G.~Liu, R.~Mawhinney and L.~Wu
for their support on the staggered $\epsilon'/\epsilon$ project.

\end{document}